\documentclass[a4paper]{article}

\usepackage[a4paper,top=3cm,bottom=2cm,left=2.5cm,right=2.5cm,marginparwidth=1.75cm]{geometry}
\usepackage[caption=false,font=footnotesize]{subfig}
\usepackage{times}
\usepackage{amssymb,amsmath}
\usepackage{graphicx}
\usepackage[lined,ruled,commentsnumbered,linesnumbered]{algorithm2e}
\usepackage{color}
\usepackage{url}
\usepackage{breakurl}

\usepackage{placeins}
\usepackage{etoolbox}
\usepackage{comment}
\usepackage{blindtext}
\usepackage{multirow}
\usepackage{scrextend}
\addtokomafont{labelinglabel}{\sffamily}
\usepackage[official]{eurosym}
\usepackage{caption}
\usepackage{float}
\usepackage{color}
\usepackage{cite}
\usepackage{amssymb,amsmath}
\usepackage{graphicx}
\usepackage[lined,ruled,commentsnumbered,linesnumbered]{algorithm2e}
\usepackage{color}
\usepackage{url}
\usepackage{placeins}
\usepackage{etoolbox}
\usepackage{comment}
\usepackage{blindtext}
\usepackage{multirow}
\usepackage{scrextend}
\addtokomafont{labelinglabel}{\sffamily}
\usepackage[official]{eurosym}
\usepackage{cite}

\begin{document}
\pagenumbering{gobble}


\title{Frictionless Authentication Systems: Emerging Trends, Research Challenges and Opportunities}

\author{Tim~Van~hamme, Vera~Rimmer, Davy~Preuveneers, Wouter~Joosen, \\ Mustafa~A.~Mustafa, Aysajan~Abidin, and Enrique~Argones~R\'{u}a
  \thanks{This work was partially supported by the Research Council KU Leuven: C16/15/058. In addition, it was also funded by FWO through SBO SPITE S002417N and imec through ICON DiskMan. DiskMan is a project realized in collaboration with imec. Project partners are Sony, IS4U and Televic Conference, with project support from VLAIO (Flanders Innovation and Entrepreneurship).}
   \thanks{T. Van hamme, V. Rimmer, D. Preuveneers, and W. Joosen are with the imec-DistriNet research group, KU Leuven, Belgium. e-mail: (\{tim.vanhamme, vera.rimmer, davy.preuveneers, wouter.joosen\}@cs.kuleuven.be); M.A. Mustafa, A. Abidin and E. Argones R\'{u}a are with the imec-COSIC research group, Departement of Electrical Engineering (ESAT), KU Leuven, Belgium. e-mail: (\{mustafa.mustafa, aysajan.abidin, enrique.argonesrua\}@esat.kuleuven.be).}
}
\date{\vspace{-5ex}}
\maketitle

\maketitle

\begin{abstract}
Authentication and authorization are critical security layers to protect a wide range of online systems, services and content. However, the increased prevalence of wearable and mobile devices, the expectations of a frictionless experience and the diverse user environments will challenge the way users are authenticated. Consumers demand secure and privacy-aware access from any device, whenever and wherever they are, without any obstacles. This paper reviews emerging trends and challenges with frictionless authentication systems and identifies opportunities for further research related to the enrollment of users, the usability of authentication schemes, as well as security and privacy trade-offs of mobile and wearable continuous authentication systems.
\end{abstract}


\section{Introduction}
\label{Introduction}

Nowadays, the ubiquitous nature of mobile and wearable devices has allowed users to access a multitude of new applications, services and content. More and more personal related information is stored on (or accessed via) personal devices such as smart phones, which enhances users' experience and convenience, 
and creates new opportunities for both, consumers and service providers. However, such access of multitude applications via personal devices also brings new 
challenges for service providers that must now secure access from a wide variety of devices~\cite{Sagiroglu2013}. Moreover, there is a continuous growth of mobile malware and other mobile security threats. Thus, it is important these mobile devices to be equipped with reliable means of authentication and authorization.

However, usually, these mobile and wearable devices have limited computational and interaction capabilities. Furthermore, because these devices are small, light, and easy to carry, there is also an associated risk in that they are susceptible to loss and theft, and easier to break. The use of context information (such as the user's current location, his typical behavior, etc.) may also trigger privacy concerns. Moreover, due to the increased prevalence of wearable and mobile applications, users nowadays expect a frictionless customer experience, making minimum effort. Taking into account these characteristics, the way users are authenticated and granted access to a wide range of online services and content becomes more challenging.

\begin{figure}[t]
\centering
\includegraphics[width=0.85\textwidth]{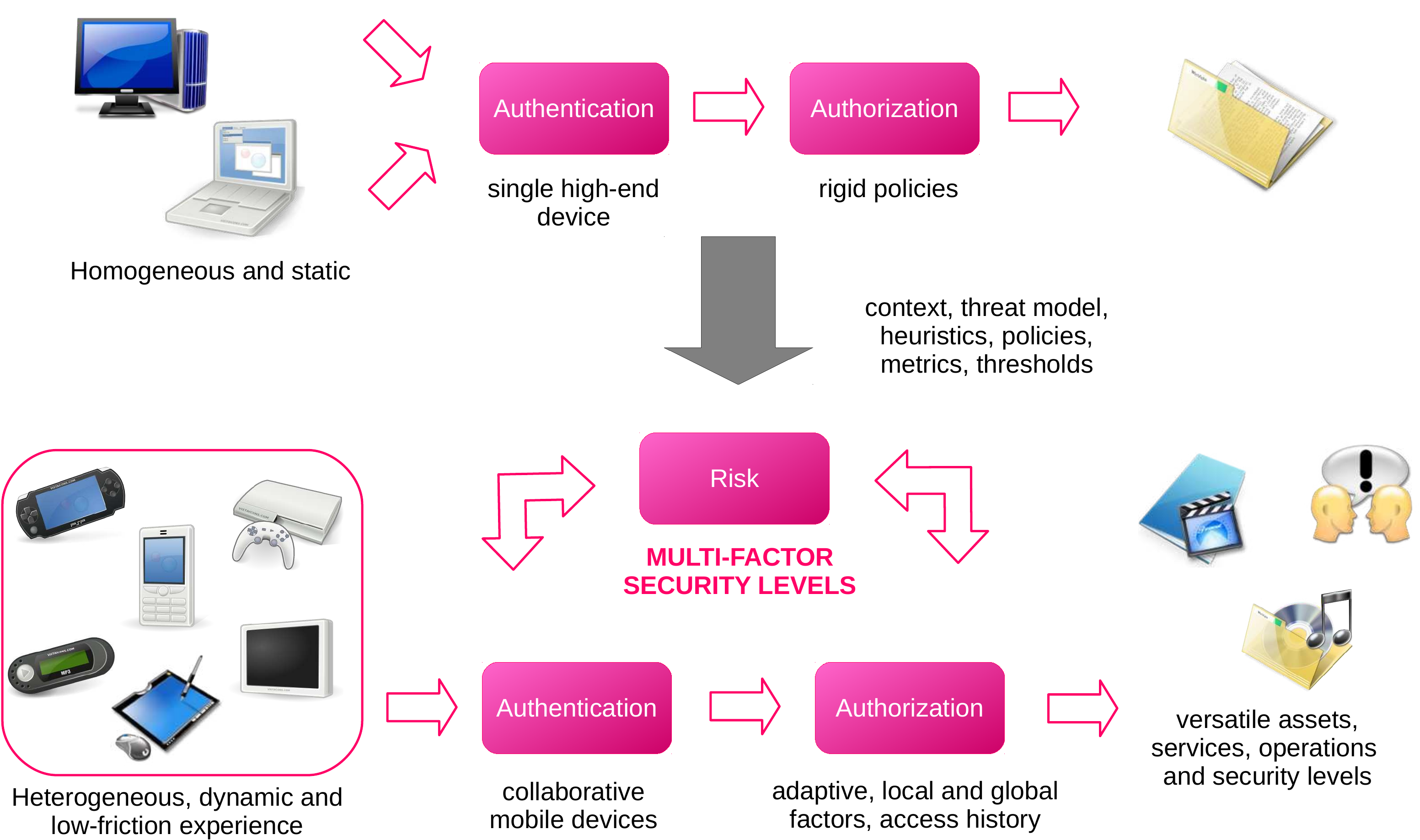}
\caption{Collaborative, frictionless and adaptive mulfi-factor authentication with many mobile devices.}
\label{fig:trends01}
\end{figure}

Ideally, users' devices will jointly and continuously operate in the background to establish the identity of the individual by continuously monitoring the context and detecting unusual deviations, as depicted in Figure~\ref{fig:trends01}. The advantage is that this will move the verification of the additional factors away from the user, making it transparent, and thereby greatly improving the convenience for the user, but posing important privacy challenges when sensitive context information is used, the addressing of which is an important aspect. The objective of pursuing a collaborative multi-device approach is that it can be less vulnerable against malicious users or unauthorized access after theft or loss of a device. Systems that support such user experience are called frictionless authentication systems~\cite{MustafaSECURWARE2017}.

In this paper we provide an overview of the emerging trends, research challenges and opportunities in such frictionless authentication systems that allow users to authenticate themselves using their devices to service providers without intentionally performing any specific authentication-related actions, such as entering a password.

The rest of this paper is structured as follows. In Section~\ref{sec:stateofpractice}, we review the current state of practice in mobile and multi-factor authentication, as well as risk-adaptive solutions. Emerging trends on collaborative and behavioral are highlighted in Section~\ref{sec:emerging}. Section~\ref{sec:challenges} reviews challenges and opportunities for further research. We conclude the paper in Section~\ref{sec:conclusions}.

\section{State-of-Practice in Authentication}
\label{sec:stateofpractice}
Before highlighting emerging trends in frictionless authentication systems, we will briefly review current best practices and the state-of-the-art in multi-factor authentication.

\subsection{Mobile and Multi-Factor Authentication}
Weak passwords are a major cause of data and security breaches~\cite{Jakobsson2013}. With dictionary attacks and optimized password cracking tools, users with simple or short (i.e., less than 8 characters) passwords are easy prey, especially if they use the same password for various services. Additionally, complex passwords are difficult to enter on mobile and wearable devices. This illustrates the generally acknowledged conception that passwords are problematic. Therefore, efforts are ongoing to replace password-based authentication with better alternatives~\cite{Bhargav-Spantzel2006,Bonneau2012,Grosse2013,Guidorizzi2013}. With multi-factor authentication, users authenticate with a combination of authentication factors, i.e., knowledge, intrinsic (biometrics) and possession. Biometric factors like speaker recognition, fingerprints, iris or retina scans cannot be forgotten, but may require expensive equipment to implement. Furthermore, such solutions require storing biometric templates, which can also be compromised and which are often cumbersome to revoke.

An interesting alternative to multi-factor mobile authentication is the Pico, a concept introduced by Stajano~\cite{Stajano2011}. The Pico is a dedicated 
hardware token to authenticate the user to a myriad of remote servers; it is designed to be very secure while remaining quasi-effortless for users. The authentication process is based on the use of public-key cryptography and certificates, making common attacks on passwords (such as sniffing, phishing, guessing, and social engineering) impossible. Although being an interesting proposal, an actual implementation is currently lacking. 

\begin{figure*}[t]
\centering
\includegraphics[width=0.85\textwidth]{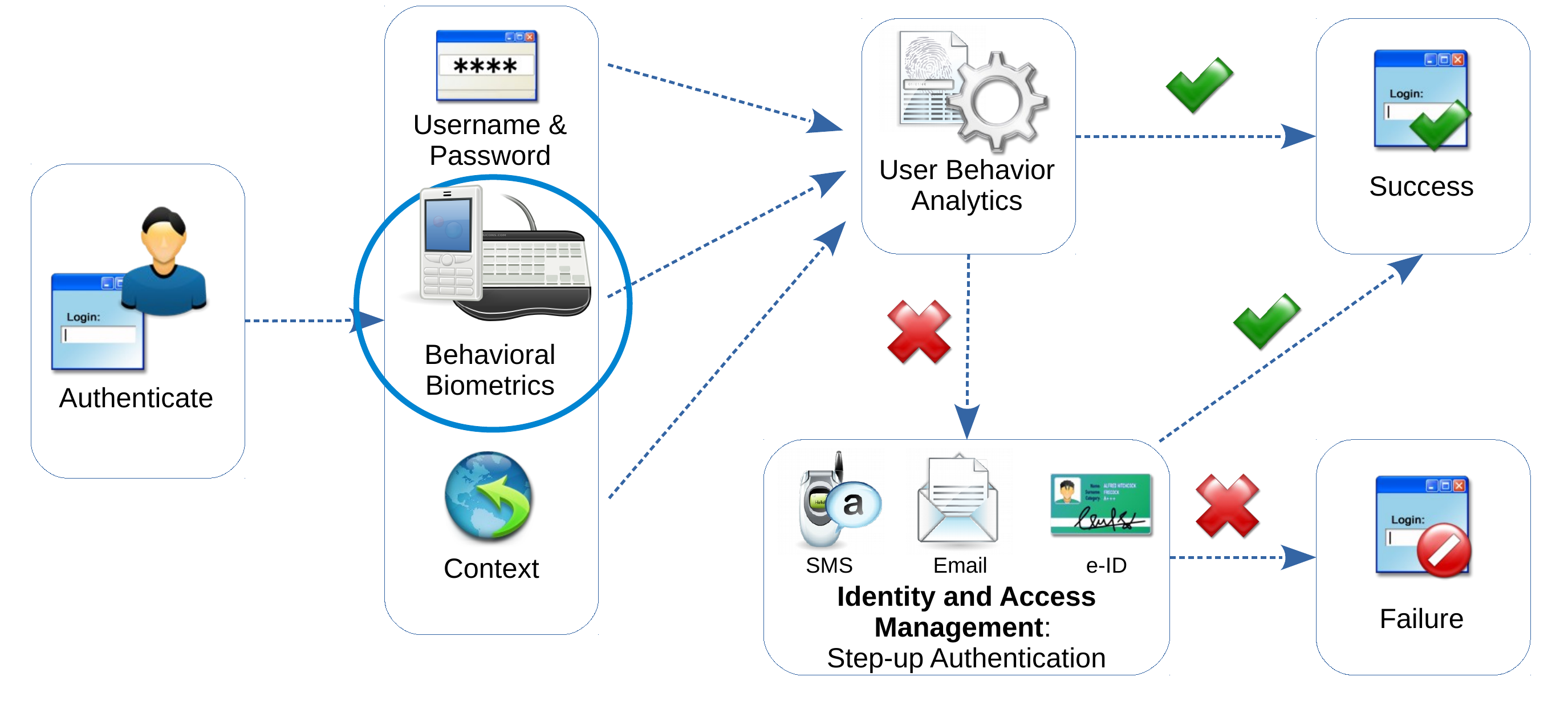}
\caption{Risk-adaptive step-up authentication leveraging context and behaviometrics adopted within contemporary Identity and Access Management systems.}
\label{fig:riskauthn}
\end{figure*}

Leveraging on these recent initiatives, dynamic, multi-factor, collaborative and context-based authentication could further improve the current state-of-the-art on mobile authentication, finding an optimal balance between cost, user-convenience and security and privacy. Early work in this direction was presented in~\cite{Preuveneers2015} in which the authors presented SmartAuth, a scalable context-aware authentication framework built on top of OpenAM, a state-of-practice Identity and Access Management (IAM) suite (see Figure~\ref{fig:riskauthn}). It uses adaptive and dynamic context fingerprinting based on Hoeffding trees~\cite{Domingos:2000:MHD:347090.347107} to continuously ascertain the authenticity of a user's identity.

However, existing solutions that exploit context information often depend on a single device. Especially for mobile devices, a simple device or browser fingerprint is hardly unique and can easily be intercepted and spoofed by an attacker~\cite{Spooren2015}.

\subsection{Risk-based Access Control and Enabling Technologies}

Authentication is a basic building block of practically all business models. As mobile devices and wearables continue to proliferate and become part of the 
user's expanded computing environment - fundamentally changing the way people access services and content - there is an associated security risk in that these devices are susceptible to loss and theft because they are small, light, and easy to carry. 

The latest trend in access control models is Risk-Adaptive Access Control (RAdAC) where access decisions depend on dynamic risk assessments. There is a 
large body of knowledge on this topic in the scientific literature~\cite{Li2013,Molloy:2012,SHAIKH2012447,Ni:2010,Santos2014,
BARACALDO2013237,KHAMBHAMMETTU201386,Kandala2011}, and risk-based authentication and access controls are being adopted in contemporary identity and access 
management solutions, such as SecureAuth IdP 8.0, RSA SecurID Risk-Based Authentication, CA Technologies and ForgeRock's OpenAM 14. Contextual 
information (device fingerprints, user location, time zone, IP address, time of day and other parameters) is used to evaluate the risk of users attempting to 
access a resource, but the approach is often based on weighted score functions or meaningless user-defined risk thresholds.

\section{Emerging Trends}
\label{sec:emerging}


\subsection{Collaborative Authentication}
Authentication means solely based on possession factors bear the risk that the unique possession factor could be lost or stolen, hence compromising the security of the authentication system. Combining these schemes with other authentication factors, such as passwords or PINs, could improve the security, but at the cost of user-friendliness. Furthermore, one still needs to take into account the typical attacks on knowledge-based authentication factors, such as 
PIN guessing or phishing attacks.  An interesting alternative are collaborative authentication schemes, where multiple devices jointly authenticate to a remote server or within a device-to-device setting. To limit the cost, the combination of wearables and the user's smartphone would be preferred. Such collaborative authentication schemes overcome the security problems of using a single possession factor during the authentication process as an adversary would have to steal multiple wearables to successfully impersonate a user, while still offering user-friendliness. Moreover, by using wearables the user is carrying anyhow, one avoids the need of employing external hardware authentication tokens, which could be quite costly.

The concept of collaborative authentication is to transform a challenge-response protocol with a single prover and verifier, to a challenge-response protocol with multiple collaborating provers and a single verifier. To mitigate the threat of wearables being stolen or lost, and the fact that the set of wearables is dynamic (the user is not always carrying the same set of wearables), threshold-based cryptography is used. The aim of threshold cryptography is to protect a key by sharing it amongst a number of entities in such a way that only a subset of minimal size, namely a threshold $t+1$, can use the key. No information about the key can be learnt from $t$ or less shares. Shamir~\cite{Sha79} was the first to introduce this concept of secret sharing. Feldman~\cite{Feldman1987} extended this concept by introducing verifiable secret sharing. Pedersen~\cite{Pedersen92} then used this idea to construct the first Distributed Key Generation (DKG) protocol. Shoup~\cite{shoup2000} showed how signature schemes such as RSA could be transformed into a threshold-based variant. 

To increase the resilience in a threshold-based authentication scheme, the number of devices included in the threshold scheme should be maximized. Therefore, Simoens~et~al.~\cite{Simoens2010} presented a new DKG protocol and demonstrated how this allows wearables not capable of securely storing secret shares to be incorporated. Peeters~et~al.~\cite{Peeters2012} used this idea to propose a threshold-based distance bounding protocol. A gap that remains to be filled is a threshold-based mobile authentication scheme, where the secret keying material is distributed among a set of personal wearables. For recent developments in continuous authentication, we refer the reader to~\cite{patel2016continuous}.

\subsection{Behaviometrics}
A recent trend in the area of continuous authentication is the use of behaviometrics. DARPA hosted the Active Authentication 
program~\cite{guidorizzi2013security} in which various kinds of behavioral biometrics, i.e., metrics that measure human behavior to recognize or verify the identity of a person, are investigated. Several studies have investigated the application of using behaviometrics in order to provide an authentication method that is (a) \textit{continuous}, during an entire user session, and (b) \textit{non-intrusive}, since the normal user interaction with the system is 
analyzed.  It has been demonstrated that a user identity can be recognized and verified by means of several behaviometrics, such as keystroke dynamics, mouse movements (together with display resolution)~\cite{de2013mouse}, gait analysis~\cite{DBLP:conf/dbsec/hammePJ17}, CPU and RAM usage~\cite{deutschmann2013continuous}, accelerometer~\cite{DBLP:conf/essos/GoethemSPJ16} and battery fingerprints of mobile devices~\cite{DBLP:journals/mis/SpoorenPJ17}, stylometry~\cite{calix2008stylometry}, web browsing behavior~\cite{abramson2013user}, etc.  An overview of techniques can be found in these works~\cite{KARNAN20111565,Deutschmann2013,Saevanee2012} and survey~\cite{wang2009behavioral}. A key challenge will be to investigate which 
combination of behaviometrics will deliver a sufficient low number of false positives (mistakenly granted access = security concern) and false negatives 
(mistakenly denied access = user experience concern) such that the risk is acceptable given the circumstances.

\section{Challenges and Opportunities}
\label{sec:challenges}

\begin{figure}
\centering
\includegraphics[width=0.40\columnwidth]{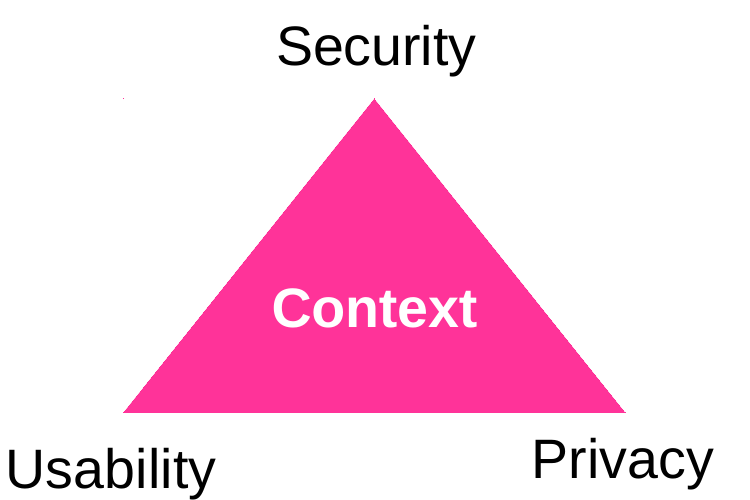}
\caption{Security, privacy and usability trade-offs in frictionless authentication.}
\label{fig:tradeoff}
\end{figure}

A frictionless authentication system is a complex system, involving multiple devices and sensors that interact with each other. This complexity makes such systems also a very flexible kind of authentication system. Nonetheless, several challenges and research opportunities remain. Authentication systems are usually characterized by the following interacting dimensions (see Figure~\ref{fig:tradeoff}):
\begin{itemize}
\item[-] \textit{Security}, which refers to how difficult it is for an impostor to be falsely authenticated.
\item[-] \textit{Usability}, which describes how easy and convenient it is for genuine users to be authenticated.
\item[-] \textit{Privacy}, which describes how any private information about the user being used are securely stored and/or processed by the system.
\end{itemize}

Security and usability are usually a trade-off in most authentication systems. For instance, False Acceptance and False Rejection Rates (FAR and FRR, respectively) are usually depicted in a ROC curve in biometric systems, and the lower the FAR is the higher the FRR is, where FAR is related to security, and FRR is related to usability. Hence, authentication systems are characterized by a specific security-usability trade-off. Regarding privacy, it can be also related to the security and usability of an authentication system. For instance, biometric systems based on protected templates, with a superior privacy protection when compared to their unprotected counterparts, usually provide an inferior set of working points regarding usability and security. In addition, the disclosure of a biometric template can lead to a security problem, unless appropriate revocation mechanisms are incorporated.

Active authentication systems involve multiple devices and sensors that interact with each other. This complexity also makes a frictionless authentication system a very flexible and powerful kind of system, which can be dynamically adapted to different usage scenarios, security-usability trade-offs, and overcome situations in which other types of authentication mechanisms would normally fail. In what follows, we expose different challenges and opportunities related to these three dimensions, \textit{security}, \textit{usability} and \textit{privacy}, and specific to frictionless authentication systems

\subsection{Security}
Regarding security, active authentication systems based on multiple behaviometrics and/or biometrics can provide increased security, since they are intrinsically multi-factor, and each employed behavioural modality makes them more difficult to spoof. However, the authentication decision will be based on the outcome of the classification and/or clustering algorithms. Such algorithms are usually not 100\% accurate~\cite{wang2009behavioral}, and in some cases the templates must be retrained by discarding old data to account for changes in the user's behaviour. This creates an opportunity for an attacker to impersonate a legitimate user by manipulating input data to compromise the learning process (i.e., a poisoning attack). 

A specific security concern in continuous authentication systems is related to the enrollment. The enrollment phase establishes the identity of the subject within the authentication systems. Typically, this is based on credentials or certificates. However, with behavioral and context-dependent authentication, the enrollment phase becomes far more challenging, especially when using a collaborative authentication relying on multiple mobile and wearable devices. In the case of other biometrics, this can be done by ensuring the identity of the user during the enrollment phase by other means. However, since the enrollment in behaviometrics is done in an uncontrolled environment, the enrollment can also pose a threat to security, since it may be easier to inject artificial data to the system. Furthermore, behavioral authentication systems relying on machine learning methods require a time-consuming training step on an individual basis before they become effective.

\subsection{Usability}
Regarding usability, the frictionless nature of continuous authentication makes these systems one of the most convenient and easy to use modalities, since the user does not even need to learn how to use the authentication system, and the authentication process is transparent, potentially providing a smooth user experience. Furthermore, the availability of different sensors and modalities opens the opportunity to provide a very flexible authentication mechanism, where the system can implement different security/usability trade-offs for controlling the access to different functionalities or services. However, this also poses a challenge regarding the design of template protection techniques, since this flexibility may increase significantly the complexity of the system. 

\subsection{Privacy}
Another key challenge with frictionless authentication systems is addressing the privacy concerns which arise when user behaviour analytics on sensitive data is used to continuously authenticate against online services. \textit{Honest but curious} service providers can use the keystrokes $-$ collected for behavioral authentication purposes $-$ to reconstruct the original text typed by the users. In addition, accelerometer data could be used by the same kind of adversary to reconstruct the whole history of a user's location. Furthermore, continuous authentication can also use physiological biometric measurements, whose implications regarding privacy are well known. Hence, employing the adequate biometric template protection mechanisms and appropriately imposing data minimality principles in the system design is even more important in continuous authentication.

\section{Conclusions}
\label{sec:conclusions}
There is a continuous quest for stronger authentication systems that at the same time offer a frictionless experience towards users of mobile and wearable devices. Context and behavioral information are nowadays being adopted in the enterprise marketplace as part of an adaptive authentication strategy that better serves the needs of the mobile consumer in diverse situational circumstances. However, irrespective of the technological advances to have multiple mobile and wearable devices collaborate to authenticate a user, the adoption of frictionless authentication will only be successful when the right balance 
between usability, security and privacy can be found that meets the demands of a diverse set of users.

\bibliographystyle{IEEEtran}
\bibliography{FrictAuthBiblio}

\end{document}